# Lithium Ion Battery Electrode Manufacturing Model Accounting for 3D Realistic Shapes of Active Material Particles

Jiahui Xu[a,b], Alain C. Ngandjong[a,b], Chaoyue Liu[a,b], Franco M. Zanotto[a,b], Oier Arcelus[a,b], Arnaud Demortière[a,b,c], Alejandro A. Franco[a,b,c,d,*]

a. Laboratoire de Réactivité et Chimie des Solides (LRCS), UMR CNRS 7314, Université de Picardie Jules Verne, Hub de l'Energie, 15 rue Baudelocque, 80039 Amiens Cedex, France
b. Réseau sur le Stockage Electrochimique de l'Energie (RS2E), FR CNRS 3459, Hub de l'Energie, 15 rue Baudelocque, 80039 Amiens Cedex, France
c. ALISTORE-European Research Institute, FR CNRS 3104, Hub de l'Energie, 15 rue Baudelocque, 80039 Amiens Cedex, France
d. Institut Universitaire de France, 103 Boulevard Saint Michel, 75005 Paris, France

* Corresponding Author: Alejandro A. Franco, Email: alejandro.franco@u-picardie.fr

**ABSTRACT**

The demand for lithium ion batteries (LIBs) on the market has gradually risen, with production increasing every year. To meet industrial needs, the development of digital twins designed to optimize LIB manufacturing processes is essential. Here, by using $LiNi_{0.33}Co_{0.33}Mn_{0.33}O_2$ (NMC111) material as an example, we introduce the realistic particles shapes of the active material obtained from X-ray micro-computed tomography into a Coarse-Grained Molecular Dynamic physical model to simulate the slurry and its drying, and into a Discrete Element Method model able to simulate the calendering of the resulting electrode. This model enables to link the manufacturing parameters with the microstructure of the electrodes and to better observe the effect of the former on the heterogeneity of the electrodes. The results of the simulations allow us, among others, to observe the alteration of the electrode heterogeneity during the manufacturing process and the slight deformation of the secondary particles of active material.

**TOC GRAPHICS**

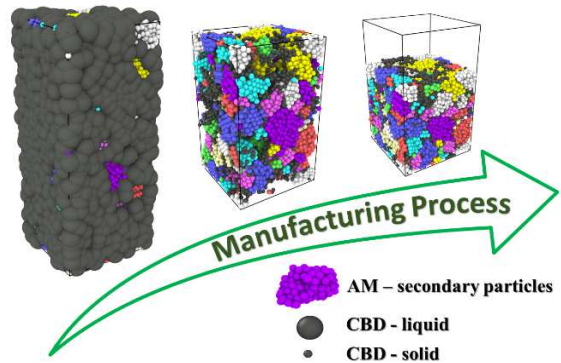

Rechargeable lithium ion batteries (LIBs) are widely used in mobile electronics, military, medical and electric public transport, and now account for a growing share of the private vehicle market[1]. In recent years, the production of LIBs has gradually increased. In response to market demands, studies have focused on achieving higher energy densities while maintaining or reducing costs[2,3].

In LIBs, the electrode is where the electrochemical reaction takes place and its complex architecture affects the rate and degree of the reaction. In order to optimize the performance of LIBs, it is essential to understand the influence of each parameter at each stage of manufacturing on the electrode architectures. In general, the properties of the electrode slurry (*e.g.* density, viscosity) will impact the coating process, which is related to the size and shape of the material, relative ratio of components and the mixing conditions[4–7]. The evaporation of the solvent affects the architecture of the dried electrode, during which the carbon binder domain (CBD) and the percolated porous structure are formed[8,9]. The calendering process, a critical step in the manufacturing process, reduces the electrode thickness increasing electrical conductivity, thermal conductivity and energy density, at the cost of reducing pore network and altering ionic conductivity[10–12]. Non-destructive techniques, such as X-ray computed tomography (XCT), are widely used to obtain the microstructure of electrodes, enabling volume-based 3D characterization. This approach has been used to study the porosity and tortuosity factor of a significant number of anode and cathode electrode active materials (AMs) like graphite, $LiNi_xMn_yCo_{1-x-y}O_2$ (NMC), $LiFePO_4$ (LFP) and $LiCoO_2$ (LCO)[13]. However, the distribution of the CBD in the electrode is limited due to the nano-features and the similarities between X-ray attenuation coefficients of carbon and pores. In order to overcome the deficiencies of neglecting the CBD, Zielke *et al.* developed a combination of X-ray tomography and a virtual design approach[15]. Lu *et al.* combined separate scans of high-attenuating NMC and low-attenuating CBD, which enabled to reconstruct the 3D electrode including microstructural heterogeneities at nanoscale[14]. Nguyen *et al.* used X-Ray holographic nano-tomography to distinguish CBD in NMC electrode[16]. In order to unveil the evolution of electrode architectures, several convolutional neural network-based image segmentation methods have been developed to improve 3D imaging and discrimination between phases[17,18].

In the previous works of our group in the context of our ARTISTIC project[19], coarse-grained molecular dynamics (CGMD) and Discrete Element Method (DEM) were used to construct the physical models of different manufacturing process steps. These steps ranged from the slurry, consisting of active materials, conductive carbon, binder and solvent[20], to the drying[21,22] and calendering[23] of the electrode by using NMC111 electrodes as an example. Likewise, models of the electrolyte infiltration[24–26] and electrochemical performance verification[23,27] were also developed to complete the digital twin. The transfer of the operational parameters of the manufacturing process into the 3D electrode microstructure makes it possible to directly relate these parameters (*e.g.* formulation, degree of compression) to the electrode microstructures (formely called by us *mesostructures*) and their associated properties (*e.g.* porosity, tortuosity factor, spatial distribution of the materials). By combining this physics-based computational workflow with machine learning and Bayesian Optimization, it is now possible to predict which manufacturing parameters to adopt in order to optimize different electrode properties simultaneously.[28,29] In the ARTISTIC models reported until now, spherical particles are used to represent the active material and the CBD, which may differ from real electrodes. In another approach, Nikpour et al.[30] generated a multi-phase smoothed particles model, simulating the drying and calendering process by using non-spherical AM particles. However, the particle morphology is generated from 2D SEM images, which limits the representativity of the real particle shapes in 3D. Furthermore, works have reported that aspherical particles lead to nonuniform lithiation during electrochemical reactions[31].

Here, we report a significant upgrade of our ARTISTIC project CGMD models to account, for the first time, for the real morphology of the active material (AM) particles. For demonstration purposes, we use here the morphology of AM particles extracted from µ-XCT characterization of NMC111-based electrodes. The 3D volume of the NMC secondary particles is used to generate the input in our new electrode manufacturing modeling workflow. The original spherical particles are replaced here by the AM with realistic shapes formed by spherical primary particles. This brings our model closer to the real electrode morphology, making it possible to capture the deformation of AM particles during calendering. The objective of this work is to introduce the realistic particles into our pre-existing manufacturing simulation workflow and to study the changes in electrode microstructure upon calendering, with a 1-to-1 comparison with µ-XCT results.

The µ-XCT measurements were performed on laboratory-prepared NMC uncalendered electrodes at 25 keV[32], as described in a former paper from our group[24]. After reconstruction, pretreatment and segmentation[33], the tomographic data are transferred to a 3D microstructure containing two phases, one which represents the AM and the other which represents both pores and CBD. The resulting 3D microstructure has an effective voxel size of 0.642 µm, while the minimum diameter of the particles is 2 µm in our previous model. Thus, the voxel size is good enough compared to the particle size distribution (PSD) in our material (1-16 µm), and the input of our physical model (2-14 µm). Then, the individual NMC secondary particles are separated and labeled by a Watershed-based Object-Separate algorithm in the commercial Avizo software (Thermo Fisher)[34]. A bounding box for each of the labeled particles is then cut from the microstructure, and saved into a 'particle database' as Python serialized objects, along with some metadata, such as their volume equivalent sphere diameters. When constructing the physical model and generating the initial structure, the particles are randomly selected from the newly generated particle database by accounting for the PSD of the real electrode, until the target AM volume is reached. In order to control for the total number of particles in the physical model, preventing our CGMD simulations from becoming too costly, the original labeled particles are downsampled by taking every 2×2×2 voxel[3]. Then, the CGMD model structure is prepared by substituting the downsampled voxels by spheres (primary AM particles) of equivalent volume, with a diameter of 1.59 µm. The secondary particles are randomly rotated and displaced before being placed into the simulation box to guarantee the randomness of the particles in the initial structure. The workflow is shown in Scheme 1. Additional details related to the generation of initial structures can be found in the Supporting Information.

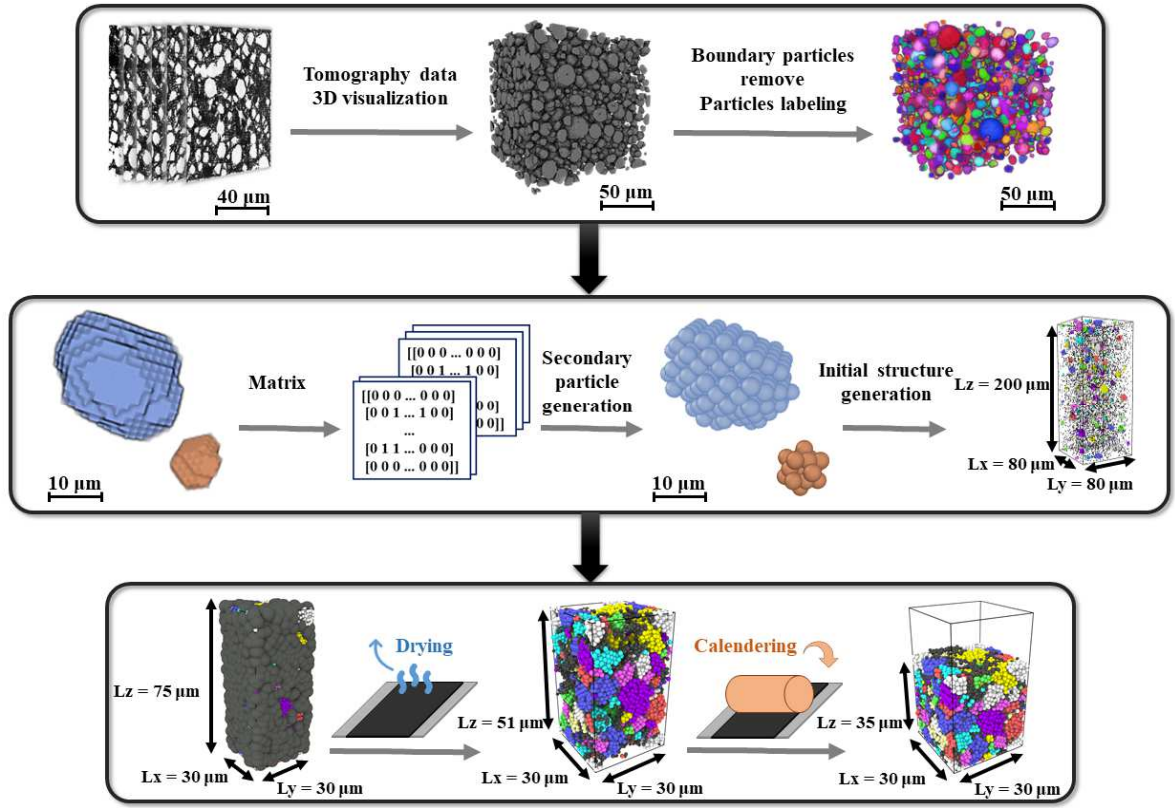

**Scheme 1**: Schematic representation of the AM structure generation procedure. A stack of 2D two-phase maps is reconstructed into a 3D volume, and the individual particles of AM are labeled. The particles cut at the border are removed. Each particle is separately transformed into a 3D matrix and stored in a database sorted by size. The secondary particles with realistic shapes are constructed from the 3D matrix by stacking spherical primary particles, and then randomly picked, rotated and displaced to generate the initial structure. The computational workflow for simulating the manufacturing process is schematized: slurry, drying and calendering simulations.

Here we simulated three steps of the manufacturing process (slurry formulation, drying and calendering) by using CGMD simulations as implemented in LAMMPS. Mixing and coating steps are not simulated explicitly in this study even though these processes can have a significant effect on the resulting 3D electrode microstructure. In the CGMD simulations, the particle beads that constitute our system can be considered as explicit particle beads, such as the primary AM particles that we generated; or as effective particles encompassing carbon, binder and solvent in the slurry model, or the effective particle encompassing carbon and binder in the dried electrode model. Those particles interact due to the action of the force fields (FFs), which have to be parameterized to mimic the experimental properties of the system[23,35,36]. In our initial microstructure, the PSD of the secondary AM particles is similar to that of the real AM particles, with a diameter of 2-14 μm. In this work, we simulate two different formulations: 94% NMC - 3% C65 – 3% PVdF (94% AM and 6% CBD in the simulation) and 96% NMC – 2% C65 – 2% PVdF (96% AM and 4% CBD). The simulation consists of 166 AM secondary particles for the 96%-4% electrode and 165 particles for the 94%-6% one. The diameter of CBD particles changes for the slurry or the dried electrode model. For the latter, particles with diameter 1.3 µm are considered, while the former includes particles of diameter 6.2 µm, with a density (0.95 g/cm$^3$) corresponding to a 50% nanoporosity. The number of CBD particles is calculated to reach the required mass according to the formulations. The FFs used in this model for the whole manufacturing process simulation include the Lennard-Jones FF (LJ) and Johnson-Kendall-Roberts model-based Granular FF (JKR), which are available in the LAMMPS software[37,38]. The real values of Young's modulus and Poisson's ratio of the material are obtained from the literature[39–41] and used as input parameters for the JKR FF. The reader is referred to the Supporting Information for a more detailed description of the CGMD procedures.

At the experimental level, µ-XCT is again used to study the changes in the geometric parameters of our *in-house* prepared electrodes, before and after calendering. The experimental procedure is described in the Supporting Information. In short, the electrode slurry is first prepared with a formulation of 96% NMC – 2% C65 – 2% PVdF, and then coated and dried over an aluminum foil. Then, the as-dried electrode is calendered until a 30% compression is obtained. Finally, the microstructures of both the calendered and uncalendared electrodes are analyzed by µ-XCT, and the obtained data is segmented into biphasic image stacks. The tortuosity factors of the microstructures are calculated using the software TauFactor[42]. It is worth pointing out that the CBD and pore phases were combined to calculate the porosity and the tortuosity factor since it is not possible to derive the morphology of the nanopores in CBD. At this point, the former is probably overestimated while the latter is probably underestimated. In addition, the sub-volumes with the same dimensions as our model are cropped out from a random location for the same study. Figure 1a shows the 3D microstructures of the uncalendered and calendered electrodes. 20 sub-volumes with the same size as our model are randomly cut out and the microstructure parameters are calculated, as shown by the hollow points presented in Figure 1b. Their average value is then compared with the value from our modeling-predicted microstructures. The sub-volumes that are most similar to the average value are selected for comparison, which are shown in Figure 1c. The volume fraction distribution of the two phases (AM and Pores+CBD) in the direction of electrode thickness (standardized) is presented in Figure 1d. From this, it is evident that the volume fraction of AM in the region near the CC and the electrode surface is lower than the average value, while the Pores+CBD shows the opposite trend. This is especially evident for the uncalendered electrode, indicating that the calendering process increases the flatness of the electrode surface.

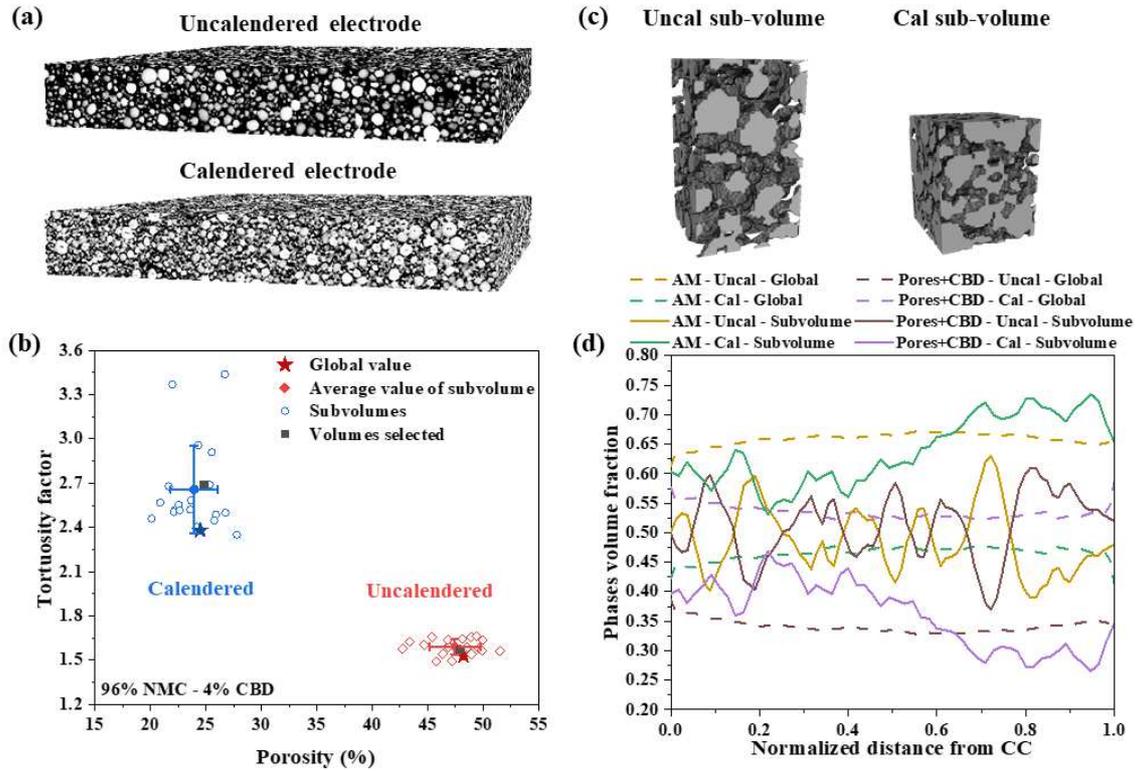

**Figure 1:** Quantitative analysis of the μ-XCT data. **a**: Global 3D microstructure of uncalendered and calendered electrodes. **b**: Subvolumes selected from uncalendered and calendered electrodes, respectively. **c**: The porosity and tortuosity factor study on the μ-XCT data. **d**: The volume fraction change of the two phases in the direction of the electrode's thickness

The computational workflow is schematized in Scheme 1. The slurry, dried electrode and calendering simulations (9494 AM particles and 5458 CBD particles for 94%-6% and 10330 AM particles and 3877 CBD particles for 96%-4%) take respectively ~24, ~11 and 9~37 h by using one node (128 GB of RAM) composed of 2 processors (Intel® Xeon® CPU E5-2680 v4 @ 2.40 GHz, 14 cores) on the MatriCS platform (Université de Picardie Jules Verne). The simulation of the slurry starts with an initial box of 80×80×120 μm$^3$ with periodic conditions in all three directions. An isothermal-isobaric (NPT) condition at 300 K and $10^5$ Pa is applied. Due to the computational limitations of LAMMPS, we were unable to achieve the viscosity simulation as in our previous works[20]. We used the experimental density values to validate the result of this step. The comparison of the simulation results and the experimental values are shown in Table 1.

**Table 1**: Comparison of experimental and simulation densities of two slurry compositions with different formulations.

| Formulation | | Density (g/cm$^3$) |
|---|---|---|
| 94% AM – 6% CBD | Experimental value | 2.03 |
| | Simulation result | 2.04 |
| 96% AM – 4% CBD | Experimental value | 2.14 |
| | Simulation result | 2.18 |

During the drying step, we consider the homogeneous evaporation model[23], where all the CBD particles of diameter 6.2 μm, containing the solvent, shrink into the solid size, with diameter 1.3 μm, instantaneously. The system is then left to equilibrate. The values of the FF parameters change during drying in order to mimic the phase changes coming from slurry (mimicking a liquid-like behavior) to the ones of the dried electrode (mimicking a solid behavior). The values of the FF parameters are reported in Table S2 of the Supporting Information. The main difference between the FF parameters of the slurry and the dried electrode is an increase in the attractive and elastic interactions, accounting for stronger particle links due to binder bridges and a greater Young modulus, respectively. In this case, instead of using periodic boundary conditions in all three directions, a fixed surface is applied in the z-direction. We calculate and define the final thickness of the dried electrode based on its experimental density and porosity and vary the thickness of the entire electrode during the non-equilibrium molecular dynamic simulation. In this way, the thickness of the electrode decreases from top to bottom (electrode surface to CC). At each time step, the temperature is rescaled in order to maintain the entire simulation at 353K (80°C). We compare the result with experimental porosity and tortuosity factor and the values obtained from the μ-XCT data for the dried electrodes. For both volumes, we considered that the CBD phase contains 50% of nanopores which is below the obtainable spatial resolution and therefore cannot be partitioned out.

Electrode calendering is another fundamental step in the battery manufacturing process. In LIBs, calendered electrodes result in increased electronic conductivities and mechanical strengths, which are necessary in order to optimize the volumetric energy and power densities of the cells. However, it comes at the risk of collapsing the electrode's pore network at too extreme compressions, considerably increasing the ionic transport resistance, which will result in lower rate-capabilities and cause incomplete utilization of the AM. Thus, it is important to analyze the effect of the calendering on the electrode microstructure, which will help choose appropriate compression parameters that prevent the abovementioned issues.

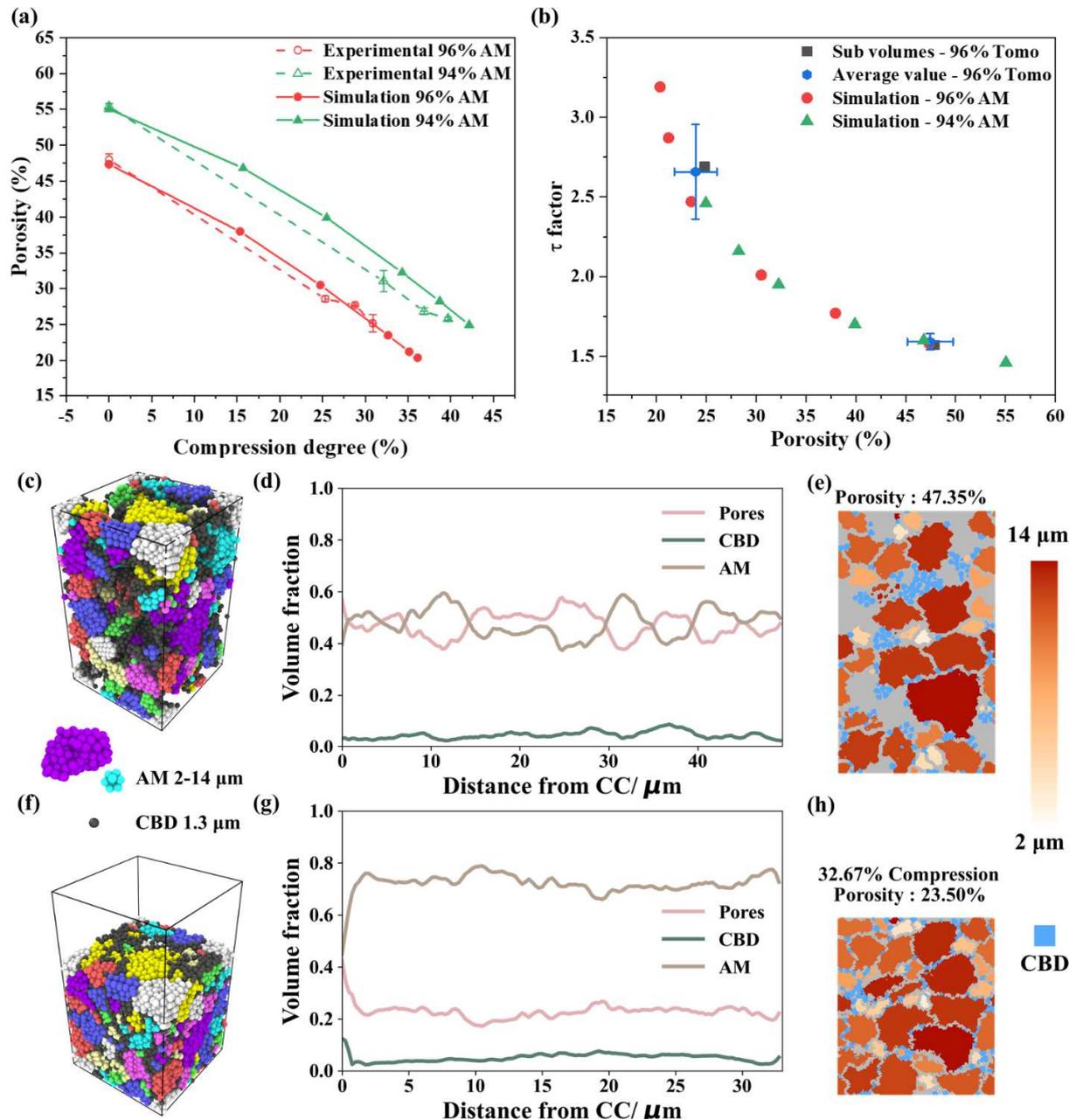

**Figure 2**: **a:** Comparison between experimental and simulated porosity for both electrode compositions as a function of compression degree. **b:** Tortuosity factor (τ) as a function of the porosity during the simulated calendering and comparison with µ-XCT data. **c and f**: The 3D structure resulting from the calendering model. **d and g**: Pore, CBD and AM volume distribution along the thickness direction, represented by black, red and blue curves respectively. **e and h**: Cross-sectional view for the electrodes. Panels c through h correspond to the 96% NMC – 4% CBD formulation. The blue represents the CBD and the red from light to dark represents different sizes of AM.

In our calendering model, a plane moves downward at a constant speed, compressing the electrode to simulate the actual process of passing the electrode through the calendering rolls. The entire process occurs at a constant 60 °C, consistent with our experiment. During the downward movement of the plane, the particles re-stabilize due to self-reorganization. After the desired degree of compression is reached, the plane is released and the electrode is "relaxed", a process in which the thickness of the electrode is partially recovered due to the elasticity of the active material. The result of the calendering process is the electrode after relaxation. Lastly, in order to eliminate the unrealistic gaps between the primary particles, while keeping the volume fraction of the AM unchanged a dilation-erosion mapping step is used as a post-processing step of the simulation results, as described in the Supporting Information.

Figures 2d-i show the distribution of the three phases along the thickness, the predicted dried and calendered electrode microstructures and the corresponding slice images. The predicted dried electrode has a 50.60%porosity while the calendered electrode has a 23.5% porosity for a 32.7% compression degree. The volume fraction of NMC particles is lower close to the CC and electrode surface, which is related to the elliptical shape of the secondary particles. Such a distribution is in agreement with the μ-XCT results of dried electrodes studied by Zhang *et al.*[9]. It is obvious that the porosity decreases and the distance between the secondary particles gradually decreases during calendering. Also, the distribution of AM in the thickness direction is more homogeneous, which can be guessed from the fact that the volume fraction (brown curves in Figure 2e and h) becomes less perturbed when the compression degree increases. The cross-sectional view is shown in Figure 2f and i, indicating the distribution of AM with different sizes. Additional information about the phase fractions for different calendering degrees is shown in the Supporting Information, which shows an increase in the flatness of the electrode surface with the degree of calendering.

Continuing with the analysis of the resulting microstructural models, we turn our attention to the porosity distribution. Here, a Watershed-based algorithm (Avizo) is used to segment the pore space into individually labelled pore-fragments. A Pore Network Model (PNM) is then used to analyze these individual pores. Here, although the pores present diverse shapes, they are approximated as spheres. An equivalent radius is calculated to represent the size and to study the size distribution. To account for the voxel size of the μ-XCT data (which represents the lower limit of our analysis), only pores with a radius greater than 1 μm are involved in the statistics. The connectivity of the pore network is evaluated by connecting neighboring pairs of individual pores with throats. The cross-sectional area of such throats is proportional to the interfacial area that the corresponding pair of pores share. Figure 3a, b, e, f, i, and j, show a comparison between the PNMs of our simulated (CGMD) and experimentally (μ-XCT) obtained microstructures for 96% AM – 4% CBD electrode. Simple visual inspection clearly shows that the simulated microstructures present more complex pore networks when compared to those of μ-XCTs, probably due to the smaller voxel sizes used to map the results from the CGMD simulations back into a voxel representation of the microstructures, which enables a higher resolution analysis of the PNM. In both cases, the calendaring process results in an overall reduction of the number of large pores, where the average radius of the pores in the simulated (experimental) calendered microstructures is reduced from 2.99 μm to 2.28 μm (from 3.76 μm to 2.16 μm), respectively. Additionally, a larger amount of smaller-sized pores appear in comparison to the uncalendered electrodes. The pore size distribution along the thickness direction and the corresponding number of pores connected to them for these four microstructures are depicted in Figure 3c, d, g and h. Upon calendering, the larger pores are reduced and replaced by smaller ones with a lower coordination number, which explains the increase in the electrode's tortuosity factor. Since the μ-XCT volumes are randomly cropped from the large electrode, this pore evolution analysis is not applicable. The cross-sectional sizes of the throats, which give an indication of the mass transfer efficiency between neighboring pores, also decrease significantly in the calendered electrodes. The quantitative results in Figure 3j indicate that narrower and more numerous channels appear after calendering.

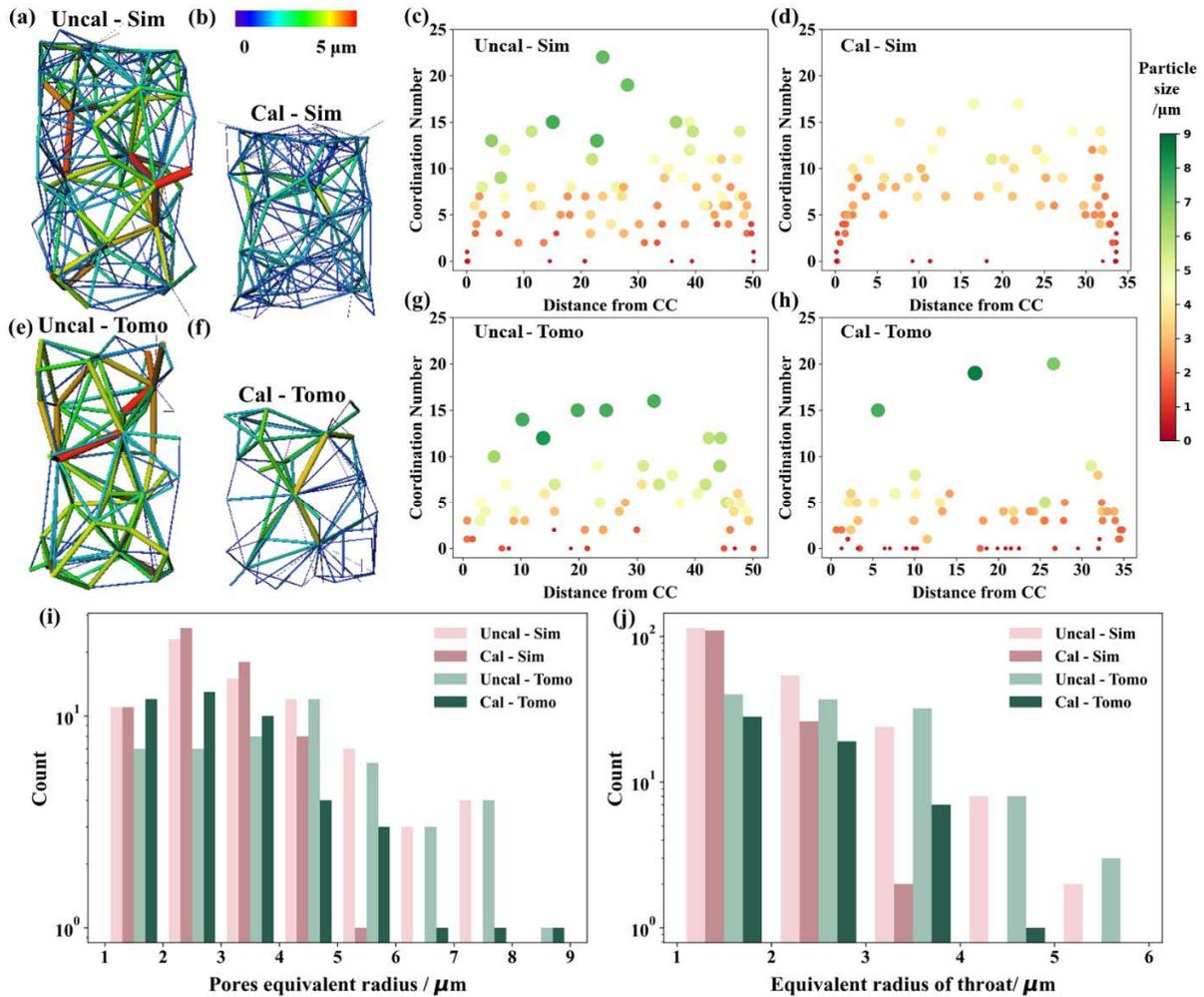

**Figure 3**: Pore network study on uncalendered and calendered electrodes from the model for 96% AM – 4% CBD electrode. **a, b, e and f:** Color-coded throats between different pores for uncalendered (a, e) and calendered (b, f) electrodes coming from simulations (a, b) and μ-XCT experiments (e, f). The color bar represents the radius of each throat. **c, d, g and h**: Size and coordination number of each pore in the PNM as a function of their position along the electrode's thickness. The color bar indicates the equivalent diameter of the pores. **i**: Global pore size distribution. **j**: Global throat size distribution.

A novel feature of our modeling approach is that it gives us the opportunity to track the deformation of individual secondary particles. The commercial software GeoDict (Math 2 Market)[43] is used to fit the individual particles into ellipses to then evaluate their Krumbein sphericity ($\sqrt[3]{bc/a^2}$, where a, b, and c are the sizes of the axes of the ellipsoid). Figure 4 gives the distribution of the Krumbein sphericities in the microstructure before and after the calendering process. It can be seen that the sphericities change due to calendering, but the peak is always around 0.7-0.8. This deformation of particles has been previously reported in the experimental literature[44]. In addition, our model offers the possibility to study the variation of particle orientation during calendering, as described in detail in the Supporting Information.

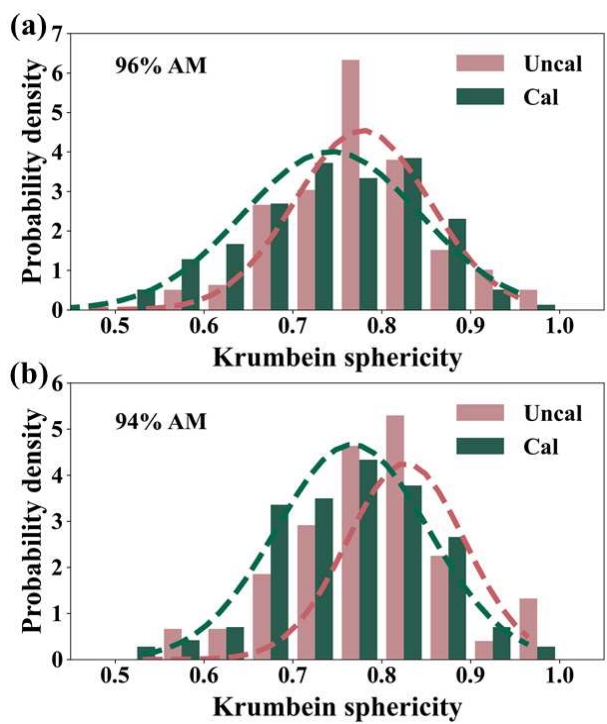

**Figure 4**: The study on deformation of secondary particles. **a and b**: The distribution of Krumbein sphericity value in the uncalendered and calendered electrode for two different formulations.

In summary, we have developed a set of models of the LIB electrode manufacturing process that uses realistic AM particle shapes. This leads to a more realistic simulation of electrode slurries, their drying and the calendering, allowing for a deeper study of the electrode heterogeneity. The experimental results of the µ-XCT of the electrode before and after the calendering are compared with the model, showing reasonable agreement. The new model captures the variation of the electrode microstructure in the manufacturing process from the particle scale, and the effect of the manufacturing parameters on the electrode heterogeneity. In addition, it makes it possible to track particle deformation and orientation changes during the calendering process. We observed that the secondary particles have suffered from small deformations under high pressure due to the mechanical properties of NMC111, and there is no obvious pattern of orientation change. This approach will be applied to other electrode materials with lower hardness, such as graphite, for further studies. This work paves the way to the simulation study of particle cracking during the calendering process which will be the subject of a future publication by us. This approach is fully compatible with the rest of our ARTISTIC online computational workflow[45], and as such, the results from this model can be used for subsequent electrolyte infiltration and electrochemical studies.

## ACKNOWLEDGMENTS

J.X., A.C.N., C. L. and A.A.F. acknowledge the European Union's Horizon 2020 research and innovation program for the funding support through the European Research Council (grant agreement 772873, "ARTISTIC" project). A.A.F. acknowledges Institut Universitaire de France for the support. F.M.Z., O.A. and A.A.F. acknowledge the European Union's Horizon 2020 research and innovation program under grant agreement No 957189 (BIG-MAP). The authors acknowledge the MatriCS HPC platform from Université de Picardie-Jules Verne for the support and for hosting and managing the ARTISTIC dedicated nodes used for the calculations reported in this manuscript. The authors acknowledge Markus Osenberg and Prof. Ingo Manke from Institute of Applied Materials, Helmholtz-Zentrum Berlin für Materialien und Energie GmbH (Germany) for the tomography data collection.